\documentclass[pra,aps,showpacks]{revtex4}
\usepackage{graphicx}
\usepackage{colordvi}
\usepackage{amsmath}
\begin{document}

\title{Interaction of solitons in dipolar Bose-Einstein
condensates \\ and formation of soliton molecules}
\author{B. B. Baizakov$^1$, S. M. Al-Marzoug$^{2,3}$, H. Bahlouli$^{2,3}$.}
\affiliation{$^1$ Physical-Technical Institute, Uzbek Academy of Sciences, 100084, Tashkent, Uzbekistan \\
$^2$ Physics Department, King Fahd University of Petroleum and Minerals, Dhahran 31261, Saudi Arabia \\
$^3$ Saudi Center for Theoretical Physics, Dhahran 31261, Saudi Arabia}
\date{\today}

\begin{abstract}
The interaction between two bright solitons in a dipolar
Bose-Einstein condensate (BEC) has been investigated aiming at
finding the regimes when they form a stable bound state, known as
soliton molecule. To study soliton interactions in BEC we employed
a method similar to that used in experimental investigation of the
interaction between solitons in optical fibers. The idea consists
in creating two solitons at some spatial separation from each
other at initial time $t_0$, and then measuring the distance
between them at a later time $t_1 > t_0$. Depending on whether the
distance between solitons has increased, decreased or remained
unchanged, compared to its initial value at $t_0$, we conclude
that soliton interaction was repulsive, attractive or neutral,
respectively. We propose an experimentally viable method for
estimating the binding energy of a soliton molecule, based on its
dissociation at critical soliton velocity. Our theoretical
analysis is based on the variational approach, which appears to be
quite accurate in describing the properties of soliton molecules
in dipolar BEC, as reflected in good agreement between the
analytical and numerical results.
\end{abstract}
\maketitle

\section{Introduction}

The interaction of solitons has been a subject of great interest
right from the beginning of their early
investigations~\cite{zabusky1965}. New fundamental features of
soliton interactions are still being discovered, in fact the
existence of a phase-dependent spatial jump in the trajectories of
two colliding matter-wave solitons, reported in a recent
experiment \cite{nguyen2014}, is just one example to be mentioned.
Apart from its scientific importance soliton interactions have
also a practical importance. For instance, interaction between
optical solitons sets the limit on the rate of information
transfer in fiber optic communication systems \cite{hasegawa1995}.
Due to important applications, soliton interactions are
extensively studied, both theoretically and experimentally, in
optical fibers
\cite{gordon1983,mitschke1987,kodama1987,mollenauer2006}, photonic
crystals \cite{stegeman1999} and plasmas \cite{lonngren1983}.
Recent experimental studies have shown that in addition to
interactions between neighboring optical solitons in close
proximity, there exists also a long range interaction between them
\cite{rotschild2006}. The generation of spatially separated
coherent matter-wave packets and their subsequent interaction
constitute the basic phenomena in the operation of modern atomic
interferometers \cite{cronin2009} working in the solitonic regime
\cite{mcdonald2014,cuevas2013} where the fringe visibility is
significantly increased compared with an ordinary atomic cloud, as
demonstrated in \cite{mcdonald2014}.

Solitons have been experimentally observed in many areas of
physics, including the Bose-Einstein condensates (BEC)
\cite{becsol,marchant2013,medley2014}. The experiments with
solitons in BEC reported so far were concerned with the creation
of solitons and studying their collective dynamics. Regarding the
type of interaction between matter-wave solitons a conjecture was
made from the behavior of neighboring solitons in a soliton train
\cite{alkhawaja2002}. Meanwhile, it would be interesting to
explore systematically the interaction between two matter-wave
solitons with varying spatial separation and relative phase.
Recent progress in controlled creation and manipulation of
matter-wave solitons in BEC
\cite{nguyen2014,marchant2013,medley2014} indicates that such
experiments on soliton interactions are now within the scope of
current technology. A key role belongs to a minimally destructive
polarization phase-contrast imaging technique \cite{bradley1997},
that allows to make multiple imaging of the soliton pair during a
single experimental run, as reported recently with regard to
phase-dependent collision of two matter-wave solitons
\cite{nguyen2014}. An essentially new method reported in Ref.
\cite{medley2014} for controlled (i.e. deterministic in both
soliton position and momentum) creation of matter-wave bright
solitons and soliton pairs without the use of Feschbach resonances
opens new perspectives for investigation of soliton interactions
in BEC with unprecedented accuracy.

Experimental realization of chromium BEC with long range
dipole-dipole atomic interactions \cite{griesmaier2005} has opened
new direction in the physics of ultra-cold quantum gases.
Subsequently two other species with strong dipolar interactions,
namely dysprosium \cite{lu2011} and erbium \cite{aikawa2012}, were
Bose-condensed. The principal difference of chromium condensates
from the alkali atom condensates is that, $^{52}$Cr has a large
permanent magnetic dipole moment $d = 6\,\mu_B$, where $\mu_B =
e\hbar/2m_e$ is the Bohr magneton. Since the dipole-dipole force
is proportional to the square of the magnetic moment, the dipolar
interactions in chromium condensate is a factor of 36 times
stronger than in alkali atom condensates, like $^{87}$Rb ($d = 1\,
\mu_B$). Similar arguments pertain also for other dipolar quantum
gases, $^{164}$Dy ($d = 10\,\mu_B$) and $^{168}$Er ($d =
7\,\mu_B$).

In this work we study, by means of variational approximation (VA)
and numerical simulations, the interaction between two bright
solitons in a dipolar BEC. We employ a strategy similar to that
used in the experimental investigation of the interaction forces
between fiber optic solitons \cite{mitschke1987,dmmol}. Following
that idea in numerical experiments we create two bright solitons
at some initial spatial separation, then give the pair a chance to
evolve for some period of time, and finally measure the distance
between the solitons when the evolution time has elapsed.
Depending on whether the distance between the solitons has
increased, decreased or remained unchanged, compared to its
initial value, we conclude about the type of soliton interaction
as being repulsive, attractive or neutral, respectively.

There is a qualitative difference between solitons in dipolar and
non-dipolar media. Specifically, two anti-phase solitons in
dipolar media attract each-other at large separation and repel at
short separation. Due to this property they can form stable bound
states with non-zero binding energy, whereas in non-dipolar media
they always repel and never form stable bound state. The
possibility of molecular type of interaction between solitons in
dipolar BEC moving in two neighboring wave-guides was shown in
\cite{nath2007}. The existence of stable multi-soliton structures
in 2D dipolar BEC was also reported in \cite{lashkin2007}.

Our main objective in this work is to find the conditions when two
interacting solitons in the same quasi-1D waveguide form a stable
bound state, which can be considered as a basic matter-wave
soliton molecule. When a stable bound state of two solitons has
been realized, we characterize the soliton molecule by its bond
length and binding energy. Our work distinguishes itself from
other relevant publications in that, we use the VA with a
Gauss-Hermite ansatz and analytically tractable function of
non-locality (response function), which allows to describe
essential features of soliton molecules in dipolar BEC. Moreover,
we provide detailed comparison of predictions of VA with the
results of numerical simulations of the Gross-Pitaevskii equation.

The paper is organized as follows. In the next section II we
introduce the governing equation and develop the VA for the
dynamics of soliton molecules in dipolar BEC. In Sec. III we use
an optimization procedure to find the shape of a soliton molecule
and validate the VA by comparing the analytical predictions with
the results of numerical simulations. In Sec. IV we present such
an important characteristic of a soliton molecule as its binding
energy. In Sec. V we reveal the character of soliton interactions
in dipolar BEC using the method borrowed from the field of fiber
optic solitons. In Sec. VI we summarize our findings.

\section{The governing equation and variational approach}

From the viewpoint of theoretical description, matter-wave
solitons in BEC and optical solitons in fibers are similar. The
mean field Gross-Pitaevskii equation (GPE) for the dynamics of BEC
and the nonlinear Schr\"odinger equation for propagation of
optical solitons in fibers have formal analogy. Similarity of the
basic equations has been fruitful in transferring many ideas from
nonlinear optics to the field of matter-waves \cite{anderson2003}.
In this paper we transfer one more idea, concerning soliton
interactions, from the field of fiber optics into the field of
BEC.

We shall consider the one dimensional GPE by taking into account
both the local and nonlocal nonlinearities, which account for the
usual contact interactions between atoms, and the long-range
dipole-dipole interactions \cite{cuevas,abdullaev2012}
\begin{equation} \label{gpe}
i\frac{\partial \psi}{\partial t} + \frac{1}{2} \frac{\partial^2
\psi}{\partial x^2} + q |\psi|^2 \psi + g \psi \int
\limits_{-\infty}^{+\infty} R(|x-\xi|) \, |\psi(\xi,t)|^2 d \xi=0,
\end{equation}
where $q=a_s/|a_{s0}|$ is the coefficient of contact interactions,
controlled by the atomic $s$-wave scattering length $a_s$, with
$a_{s0}$ being its background value, $g= a_d/|a_{s0}|$ is the
coefficient of nonlinearity, responsible for the long - range
dipolar atomic interactions, expressed via characteristic dipole
length $a_d=\mu_0 d^2 m/(12 \pi \hbar^2)$, with $m, d$ being the
mass and magnetic dipole moment of atoms, oriented along the $x$
axis, $\mu_0$ is the permeability of vacuum. Time and space are
expressed in units of $t_0 = \omega_{\bot}^{-1}$ and $l_0 =
\sqrt{\hbar/(m \omega_{\bot})}$, respectively, with
$\omega_{\bot}$ being the frequency of radial confinement. The
wave function is re-scaled as $\psi = \sqrt{2 |a_{s0}|}\Psi$ and
normalized to the reduced number of atoms in the condensate
$N=\int_{-\infty}^{+\infty} |\psi(x)|^2 dx$, which is a conserved
quantity of Eq. (\ref{gpe}). The following two models for the
kernel (nonlocal response functions) are relevant to dipolar
condensates confined to quasi-1D traps
\begin{eqnarray}
R_{1}(x) &=& (1+2x^{2}) \, \exp(x^2) \, \mathrm{erfc}(|x|)-2\pi^{-1/2}|x|, \label{R1} \\
R_{2}(x) &=& \delta^3 (x^2+\delta^2)^{-3/2}. \label{R2}
\end{eqnarray}
The former kernel was derived for the dipolar BEC using the single
mode approximation \cite{sinha}, while the latter containing a
cutoff parameter $\delta$, was proposed in Ref. \cite{cuevas} and is
more convenient for analytical treatment. Making use of the
matching conditions
\begin{equation}
R_{1}(0)=R_{2}(0), \qquad \mbox{and} \qquad
\int_{-\infty}^{\infty} R_1(x) dx = \int_{-\infty}^{\infty} R_2(x) dx,
\end{equation}
which requires $\delta = \pi^{-1/2}$, one can take advantage of the
simplicity of $R_{2}(x)$ for the application of VA. The meaning of
$\delta $ is the effective size of the dipole. Actually, it takes
value of the order of the transverse confinement length, which
makes the model one-dimensional, and sets the unit length in Eq.
(\ref{gpe}). Therefore, the choice of $\delta =\pi^{-1/2} \approx
0.56$ is quite reasonable. In the limit $x \gg \delta$, where
dipole-dipole interaction effects dominate the contact interaction
effects, both response functions behave as $ \sim 1/x^3$. This
justifies application of the kernel function $R_2(x)$ for
analytical treatment of dipolar effects in BEC. By comparing the
graphics of these two response functions one can be convinced,
that indeed $R_1(x)$ and $R_2(x)$ match very closely
\cite{cuevas}.

The Lagrangian density generating the Eq. (\ref{gpe}) is
\begin{equation} \label{lagden1}
{\cal L} = \frac{i}{2}(\psi \psi^{\ast}_t - \psi^{\ast}\psi_t) +
\frac{1}{2} |\psi_x|^2 - \frac{q}{2}|\psi|^4 -
\frac{g}{2}|\psi(x,t)|^2 \int \limits_{-\infty}^{\infty} R(x-\xi)
|\psi(\xi,t)|^2 d \xi.
\end{equation}
To study soliton interactions in dipolar BEC we need to develop
the VA for a two-soliton molecule. To this end we employ a
Gauss-Hermite (GH) trial function, which was successful in the
description of soliton molecules in dispersion-managed optical
fibers \cite{pare1999}
\begin{equation}\label{ansatz2}
\psi(x,t)=A(t) \, x \, \exp \left[ -\frac{x^2}{2a(t)^2} + i b(t)
x^2 + i\phi (t) \right],
\end{equation}
where $A(t)$, $a(t)$, $b(t)$ and $\phi(t)$ are variational
parameters, associated with the amplitude, width, chirp and phase,
respectively. The norm $N = \int |\psi(x)|^2 dx = A^2 a^3
\sqrt{\pi}/2$ is proportional to the number of atoms in the
condensate. For specified values of $A$ and $a$, the waveform
(\ref{ansatz2}) can be well approximated by two anti-phase
Gaussian functions with the amplitude $A_0$, width $a_0$ and half
separation $x_0$
\begin{equation}\label{Aax}
A_0 = \frac{2 A a}{\sqrt{\pi}} e^{-2/\pi}, \quad a_0 = \frac{\pi a
}{16} e^{4/\pi}, \quad x_0 = \frac{2 a}{\sqrt{\pi}}.
\end{equation}
Substitution of the ansatz (\ref{ansatz2}) and response function
(\ref{R2}) into the Lagrangian density (\ref{lagden1}) and
subsequent integration over the space variable $x$ yields the
averaged Lagrangian
\begin{equation}\label{lagr}
\frac{L}{N} = \frac{3}{2}a^2 b_t + \phi_t + \frac{3}{4a^2} + 3 a^2
b^2  - \frac{3 \, q \, N }{8 \sqrt{2\pi}\, a} - \frac{3 \, g \,
\delta \, N }{8 \sqrt{2} \, a} \ \left[{\cal
U}\left(\frac{1}{2},0,z \right) - \frac{\delta^2}{3 a^2} {\cal
U}\left(\frac{3}{2},1,z \right) + \frac{\delta^4}{4 a^4} {\cal
U}\left(\frac{5}{2},2,z \right) \right],
\end{equation}
where
\begin{equation}
{\cal U}({\rm a},{\rm b},z) = \frac{1}{\Gamma({\rm
a})}\int_0^{\infty} e^{-z t} t^{{\rm a}-1} (t+1)^{{\rm b}-{\rm
a}-1}dt
\end{equation}
is the confluent hypergeometric function \cite{abramowitz}, and
$z = \delta^2/(2a^2)$.

\begin{figure}[htb]
\centerline{\includegraphics[width=8cm,height=6cm,clip]{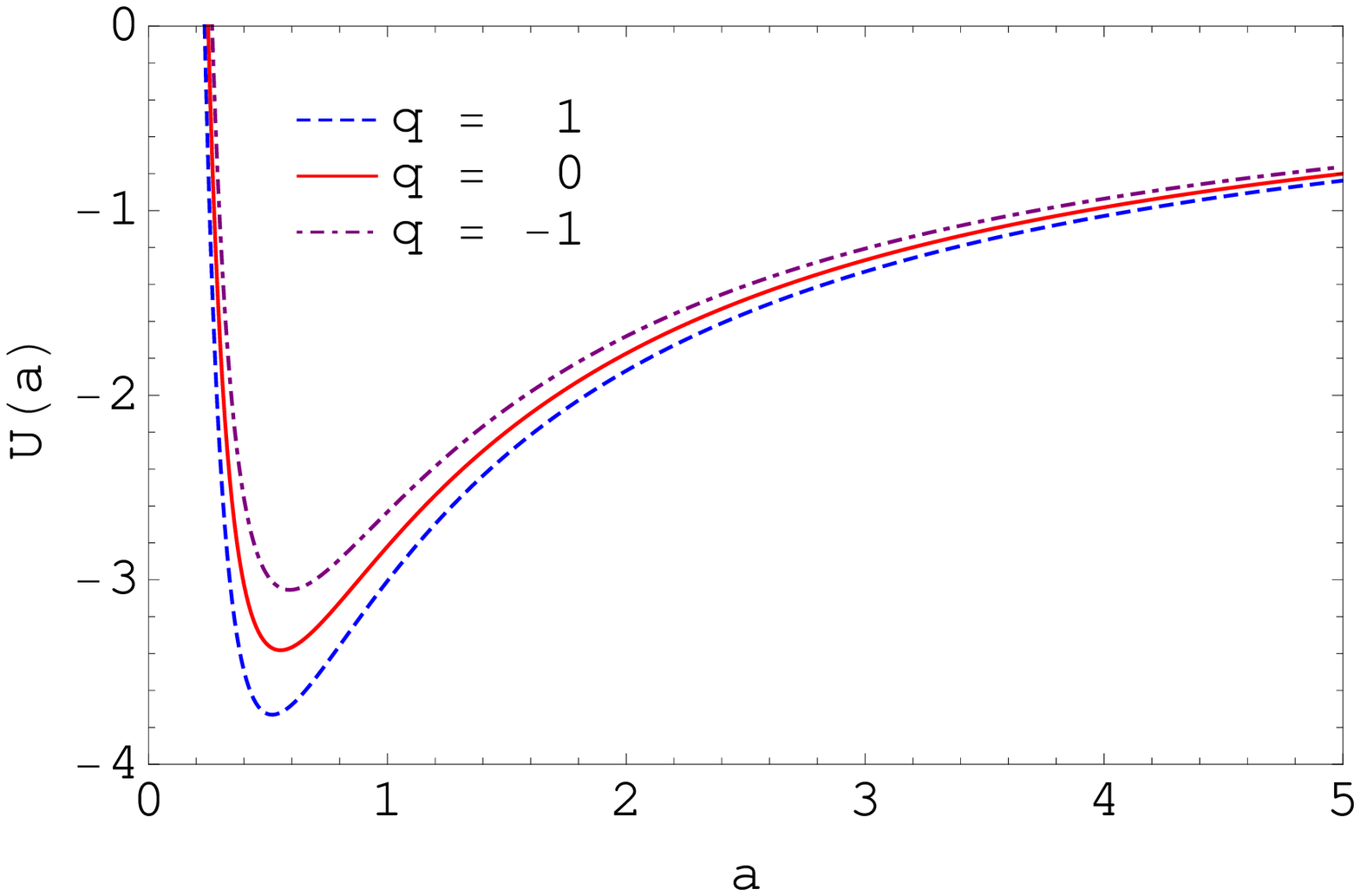}\quad
            \includegraphics[width=8cm,height=6cm,clip]{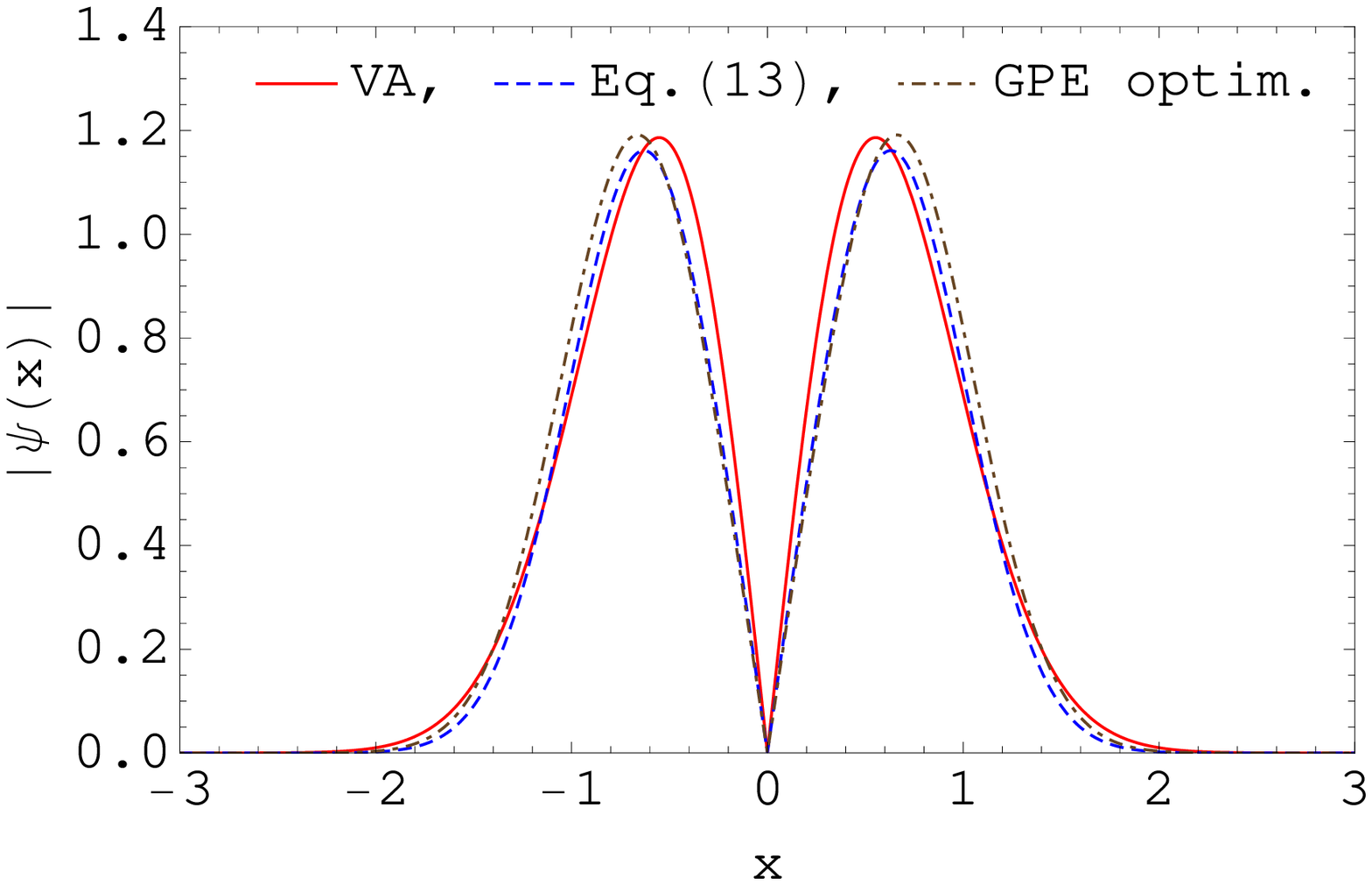}}
\caption{(Color online) Left panel: A molecular type potential
associated with VA Eq. (\ref{attmol}) for different strengths of
the contact interaction. Right panel: The shape of a two-soliton
molecule in a pure dipolar BEC with $N=2$, $q=0$, $g=20$ as
predicted by VA with trial function (\ref{ansatz2}) for $A=3.538$,
$a = 0.553$, $N=1.876$ (red solid line), by two anti-phase
Gaussian functions with parameters given in Eq. (\ref{Aax}) (blue
dashed line), and found from numerical optimization procedure,
applied to GPE (\ref{gpe}) (brown dash-dot line). The minimum of
the effective potential $U(a)$ is attained at $a_0 = 0.53$, and
the equilibrium distance between center-of-mass positions of
pulses predicted by Eq. (\ref{Aax}) is $\Delta_0 = 4
a_0/\sqrt{\pi} \simeq 1.2$, while the GPE optimization gives
$\Delta_0 \simeq 1.3$.} \label{fig1}
\end{figure}

The VA equation for the parameter $a$ of the two-soliton molecule,
which is proportional to the separation between solitons, can be
derived from the Euler-Lagrange equations $d/dt(\partial
L/\partial \nu_t) -
\partial L/\partial \nu = 0$ for variational parameters
$\nu \rightarrow a, b, \phi$, using the averaged Lagrangian
(\ref{lagr})
\begin{equation}\label{attmol}
a_{tt}  = \frac{1}{a^3} - \frac{q \, N }{4 \sqrt{2\pi}\, a^2} -
\frac{g \, \delta \, N }{4 \sqrt{2} \, a^2} \ \left[{\cal
U}\left(\frac{1}{2},0,z \right) - 3z{\cal U}\left(\frac{3}{2},1,z \right) +
7z^{2} {\cal U}\left(\frac{5}{2},2,z \right) -5z^{3} {\cal U}\left(\frac{7}{2},3,z \right) \right].
\end{equation}
The corresponding effective potential $U(a)$ for the width is
depicted in Fig. \ref{fig1} (left panel). The analytic form of the
potential $U(a)$, which can be found by integrating the right hand
side of Eq. (\ref{attmol}), is rather complicated and we do not
show it here explicitly. The fixed point $a_{tt} = -\partial
U(a)/\partial a = 0$ of this equation $a_0$ is associated with the
stationary separation between center-of-mass positions of two
solitons constituting the molecule $\Delta_0 =2 x_0 = 4
a_0/\sqrt{\pi}$. At larger separation ($a>a_0$) the solitons
attract each other ($\partial U/\partial a > 0$), and at smaller
separation ($a < a_0$) they repel ($\partial U/\partial a < 0$),
therefore the effective potential $U(a)$ has a property of
molecular type. Fig. \ref{fig1} (right panel) illustrates the
shape of a two-soliton molecule, as predicted by VA, by two
anti-phase Gaussian functions with parameters given in Eq.
(\ref{Aax}), and by optimization procedure, applied to GPE
(\ref{gpe}), described in the next subsection.

\begin{figure}[htb]
\centerline{\includegraphics[width=5.5cm,height=6cm,clip]{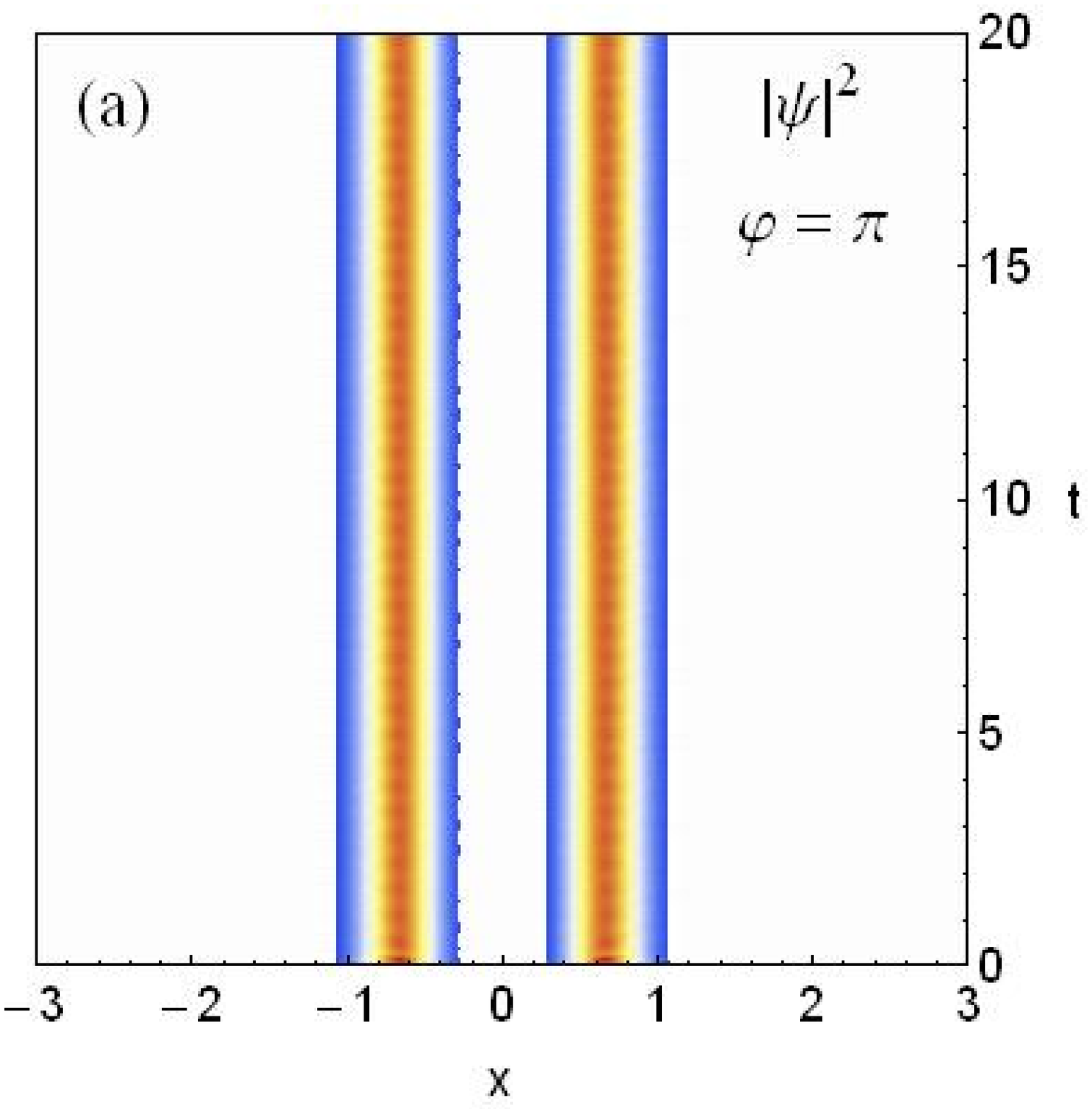}\quad
            \includegraphics[width=5.5cm,height=6cm,clip]{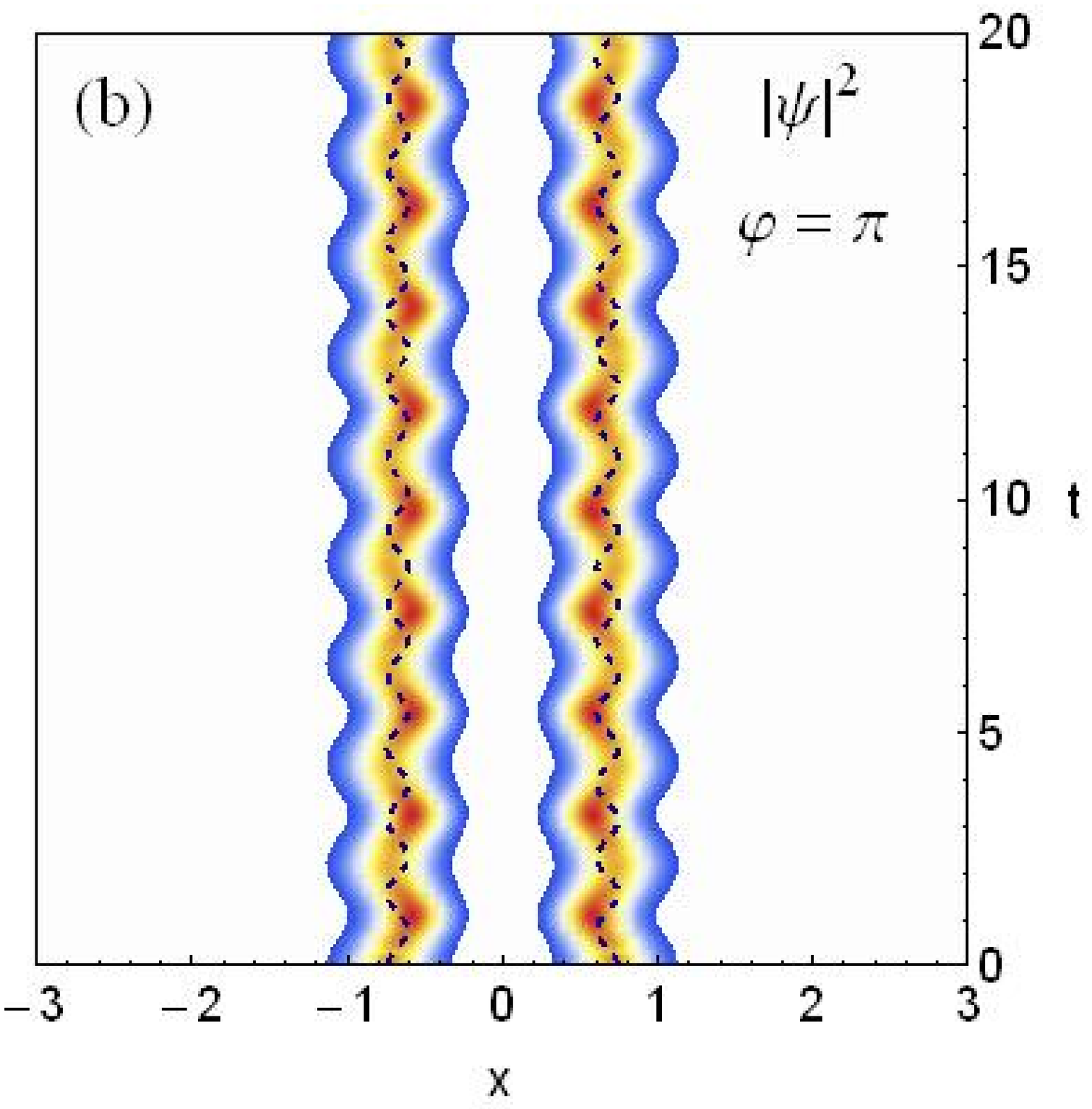}\quad
            \includegraphics[width=5.5cm,height=6cm,clip]{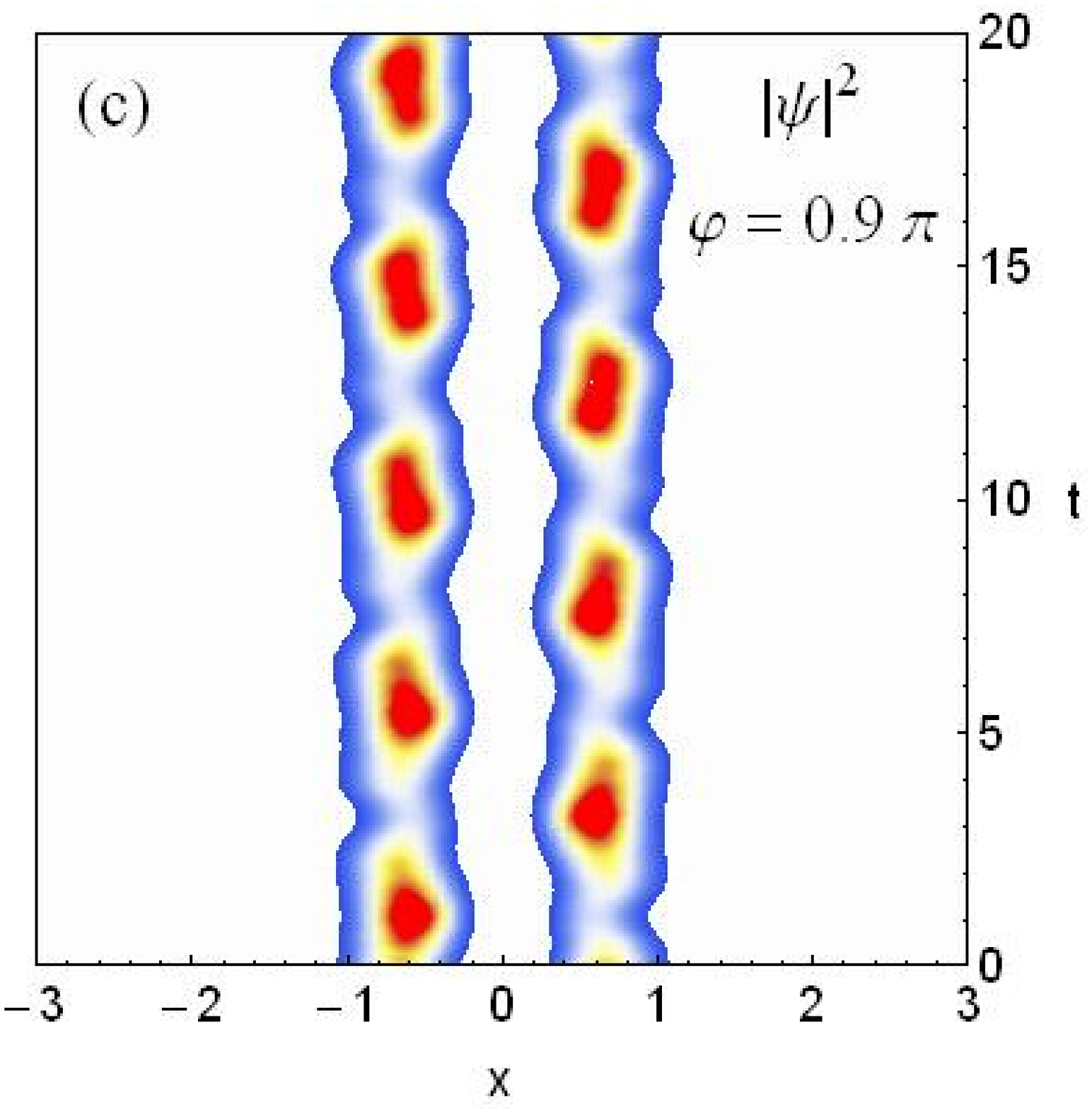}}
\caption{(Color online) (a) Stable propagation of a two-soliton
molecule composed of two anti-phase ($\varphi=\pi$) Gaussian
pulses with parameters $A_0=1.196$, $a_0=0.394$, placed at their
equilibrium positions $x_0 = \pm 0.657$. (b) When the solitons are
slightly (by 20 \%) displaced from equilibrium positions, they
perform oscillations. Density plot $|\psi|^2$ is obtained by
numerical solution of the GPE (\ref{gpe}). Dashed line corresponds
to calculations according to VA equation (\ref{attmol}), which
shows increasing phase shift with respect to GPE. (c) Periodic
exchange of atoms between two solitons when the initial phase
difference is slightly decreased.} \label{fig2}
\end{figure}

When solitons of the molecule are placed at their equilibrium
positions, they stay motionless, as shown Fig. \ref{fig2}. If
solitons are slightly displaced and released, they perform small
amplitude oscillations around their stationary separation. The
dynamics of the molecule strongly depends on the initial phase
difference between solitons. In particular, even slight deviation
from anti-phase configuration leads to periodic exchange of atoms
between solitons. At larger deviation the soliton molecule does
not form.

The frequency of soliton oscillations near equilibrium state can
be estimated from linearized version of Eq. (\ref{attmol})
\begin{eqnarray}\label{omega1}
\Omega_0^2  &=& \frac{3}{a_0^4}-\frac{N q}{2 \sqrt{2 \pi }
a_0^3}\frac{\delta  g N}{4 \sqrt{2} a_0^3} \left[-15 z_0^2
\mathcal{U}\left(\frac{5}{2},0,z_0\right)+21 z_0^2
\mathcal{U}\left(\frac{5}{2},1,z_0\right) \right. \nonumber \\
& & - \left. 9 z_0^2 \mathcal{U}\left(\frac{5}{2},2,z_0\right)+20
z_0 \mathcal{U}\left(\frac{3}{2},-1,z_0\right)-28 z_0
   \mathcal{U}\left(\frac{3}{2},0,z_0\right)+13 z_0 \mathcal{U}\left(\frac{3}{2},1,z_0\right)-2
   \mathcal{U}\left(\frac{1}{2},0,z_0\right) \right].
\end{eqnarray}
For the period of oscillations near the stationary separation we
have $T_{VA} = 2\pi/\Omega_0 \simeq 1.54$. The prediction of VA is
in qualitative agreement with the result of numerical simulation
of the GPE,  $T_{GPE} \simeq 2.2$ (see Fig. \ref{fig2}). In
general, the VA provides fairly good description of the static and
dynamic properties of the soliton molecule, while its waveform
remains close to the selected trial function (\ref{ansatz2}). The
agreement between VA and GPE deteriorates at large separation
between solitons, close to the dissociation point, when the trial
function cannot be well approximated by two anti-phase Gaussian
functions.

\section{Improving the shape of the soliton molecule by
optimization procedure}

The VA provides approximate waveform of a soliton molecule. When a
trial function with parameters, defined by stationary solution of
VA equation, is assigned as initial condition for the GPE, small
amplitude oscillations of the molecule's shape and separation
between pulses is observed. This implies that soliton molecule is
in its excited state.

For some precise parameter calculations, such as the binding energy of
soliton molecules, a truly ground state should be employed. In
Ref. \cite{golami2014} an optimization strategy to find the
stationary shape of a soliton molecule in dispersion-managed
optical fibers was proposed. Below we extend this approach for
soliton molecules in dipolar BEC. It is based on the Nelder-Mead
(NM) nonlinear optimization procedure \cite{nelder1965}, which
seeks to minimize an objective (or cost) function
\begin{equation}\label{cost}
f=\frac{1}{N_0}\int \limits^{\infty}_{-\infty}
(|\psi(x,0)|-|\psi(x,t_1)|)^2\,dx, \qquad N_0 = \int
\limits^{\infty}_{-\infty} |\psi(x,0)|^2\,dx,
\end{equation}
where
\begin{equation}\label{init}
\psi(x,0) = A_0
\left(\exp\left[-\frac{(x-x_0)^2}{2a_0^2}\right]-\exp\left[-\frac{(x+x_0)^2}{2a_0^2}\right]\right)
\end{equation}
is the initial waveform, composed of two anti-phase Gaussian
functions, separated by a distance $2\,x_0$, and $\psi(x,t_1)$ is
the result of evolution of $\psi(x,0)$ for some period of time $t
= t_1$, according to GPE. Normalization factor $N_0$ in Eq.
(\ref{cost}) is introduced to avoid trivial solutions, in
particular corresponding to $x_0=0$. In general case minimization
of the objective function can be performed with respect to
variables $a_0$ and $x_0$, since the amplitude $A_0$ is fixed by
the norm of the Gaussian. However, numerical experiments show
that, VA predicted values of $a_0$ and $A_0$ for a single soliton
are quite accurate, and minimization only with respect to pulse
separation $x_0$ can produce stationary state of the molecule. The
evolution time $t_1$ can be estimated as half period of
oscillation for the molecule $t_1 = \pi/\Omega_0$. Although the NM
optimization procedure finds the minimum of the objective function
(\ref{cost}) for Gaussian functions with broad range of
parameters, the convergence rate can be improved by selecting the
initial waveform close to the stationary state. The VA can provide
the waveform which is close to the stationary state. We find
stationary pulse separation $x_0$ and norm of the soliton molecule
$N$ from NM optimization procedure. The obtained results were
confirmed by alternative method of Luus-Jaakola
\cite{lj1973,almarzoug2006}. Our preference of these optimization
methods is motivated by several advantages, such as simplicity of
programming (since calculation of function derivatives is not
required), high convergence rate and reliability and effectiveness
in locating the global minimum of the objective function.

\section{Interaction potential and binding energy of soliton
molecules}

\begin{figure}[htb]
\centerline{\includegraphics[width=8cm,height=6cm,clip]{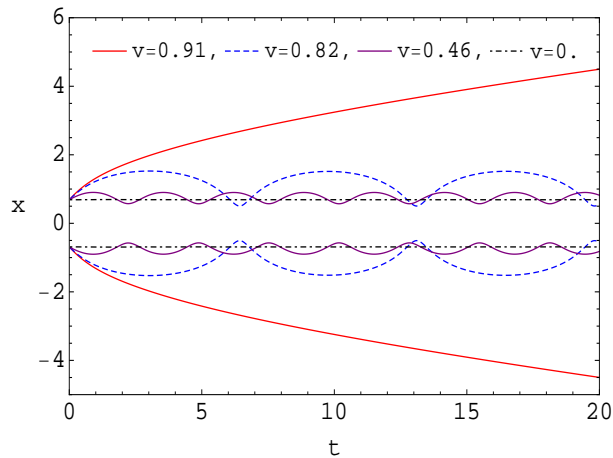}}
\caption{(Color online) Dynamics of center-of-masses of two
solitons, forming the molecule, when the two solitons are set in
motion with different velocities in opposite directions. When the
velocity is less than critical $v < v_{cr} = 0.91$, solitons
perform oscillations near equilibrium positions. At critical
velocity the molecule dissociates into freely moving individual
solitons (red solid lines). } \label{fig3}
\end{figure}

The binding energy of a soliton molecule $E_b$ can be defined as
the amount of energy, which is required for the dissociation of
the molecule into two separate individual free solitons, far away
from each other. In numerical simulations using the GPE, the
process of dissociation can be implemented by assigning an initial
velocity to each soliton in opposite directions $\psi = \psi_1
e^{ivx} + \psi_2 e^{-ivx}$. If the velocity is smaller than some
critical value $v < v_{cr}$, solitons perform oscillations around
their stationary positions, otherwise the molecule disintegrates
into individual solitons, travelling in opposite directions, as
illustrated in Fig. \ref{fig3}.

\begin{figure}[htb]
\centerline{\includegraphics[width=8cm,height=6cm,clip]{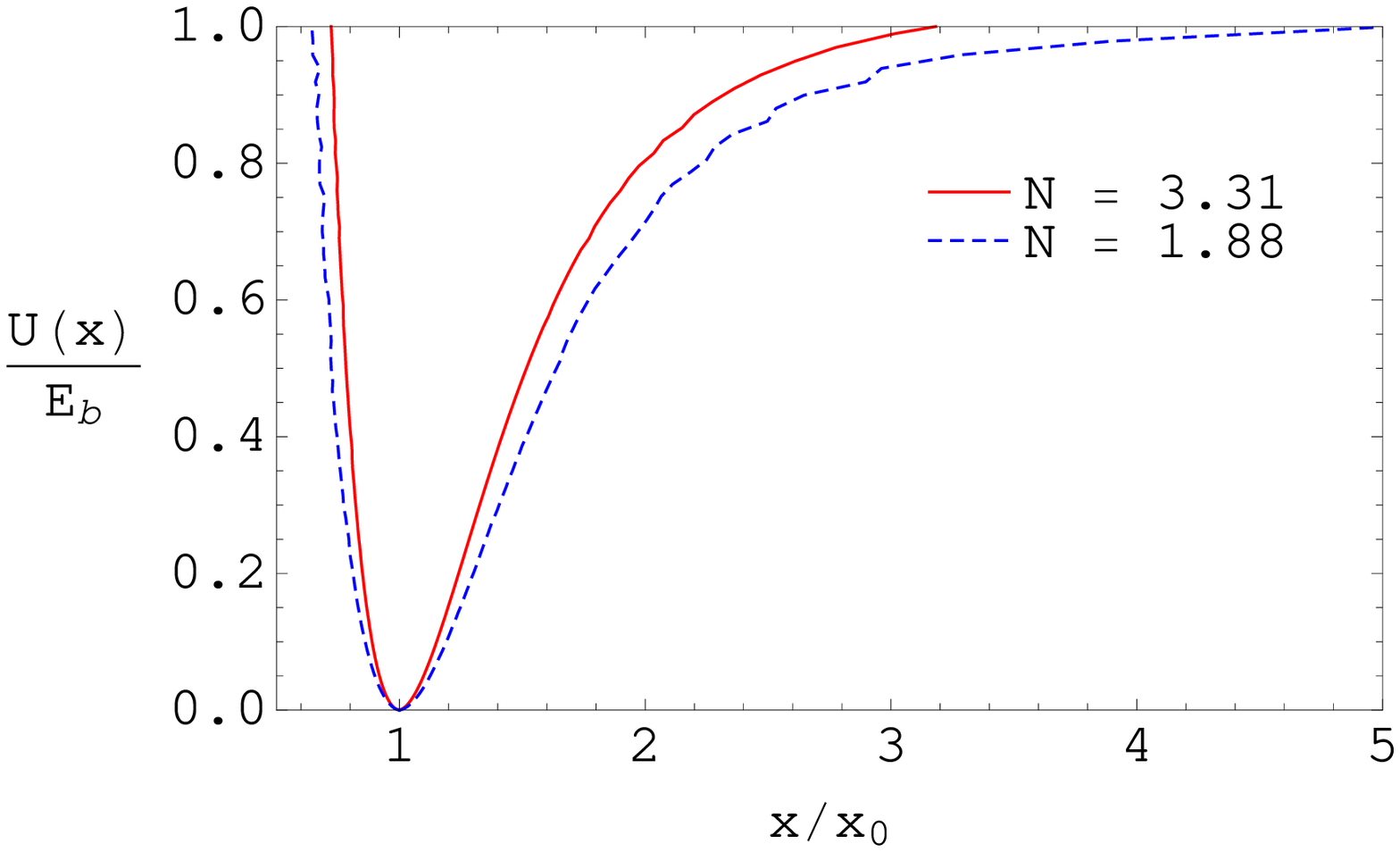}\qquad
            \includegraphics[width=8cm,height=6cm,clip]{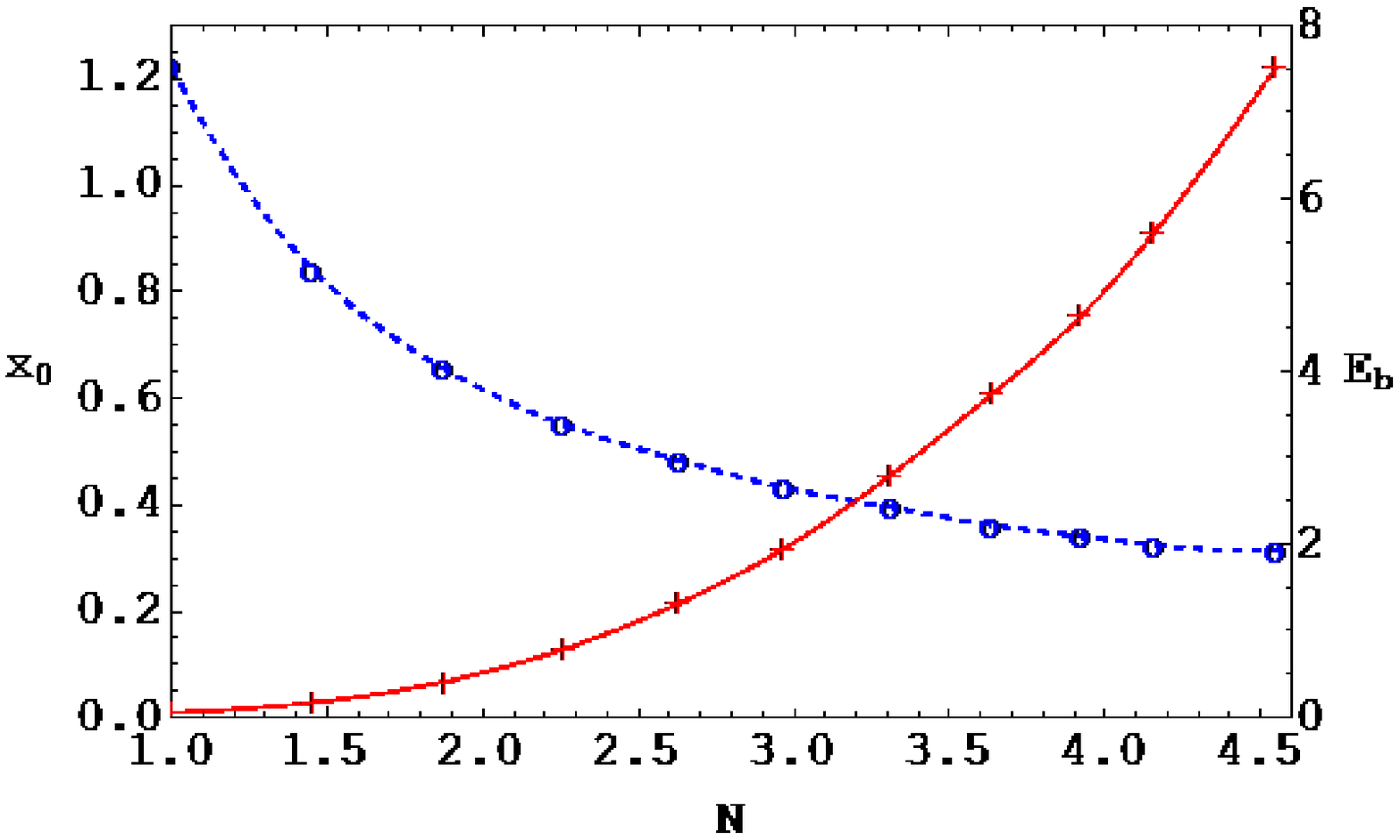}}
\caption{(Color online) Left panel: The potential of interaction
between two solitons, retrieved from numerical GPE simulation, for
two values of the molecule's norm. Similarity with the VA
predicted potential $U(a)$ in Fig. \ref{fig1} is evident. Right
panel: The stationary half separation between solitons of the
molecule $x_0$ (blue dashed line) and its binding energy $E_b$
(red solid line) as a function of the molecule's norm $N$. Symbols
correspond to values found from numerical simulations of the GPE
(\ref{gpe}), and lines are interpolating curves for visual
convenience. } \label{fig4}
\end{figure}

In a ``particle in potential well" picture this situation
corresponds to the escape of the particle from the potential well
at critical kinetic energy. The critical velocity determines the
binding energy of the molecule $E_b \sim v_{cr}^2/2$.
Fig.~\ref{fig4} illustrates the potential of interactions between
the two solitons, normalized to binding energy, as a function of
distance between solitons in units of stationary separation $x_0$.
To construct the potential $U(x)$ we assign velocity to solitons
and determine the maximal and minimal value of the separation,
which correspond to right and left classical turning points of the
oscillating particle in the potential well. Repeating these
calculations for velocities in the range $v \in [0, v_{cr}]$ we
construct the potential, shown in the left panel of Fig.
\ref{fig4}. As expected, the bigger norm $N$ (or number of atoms)
of the molecule corresponds to stronger potential, connecting
solitons.

The critical velocity $v_{cr}$, at which the molecule
disintegrates into far separated individual solitons, is
determined from GPE simulations, by setting in motion the two bound
solitons in opposite directions, as shown in Fig. \ref{fig4}. In
the experiment pushing the solitons in opposite directions can be
realized by means of a laser beam, directed into the center of the
molecule, as it was used to split the condensate in two halves
\cite{nguyen2014}. The intensity of the laser beam can be varied
to give desired initial velocity to solitons.

About the repulsive interaction between two anti-phase solitons
the following comment is appropriate. As experimentally
demonstrated in \cite{nguyen2014} and theoretically shown in
\cite{rag2015}, two colliding wave packets exchange not only
velocities (as classical particles do), but also their entire
wave-functions (as quantum mechanical particles do via tunnel
phenomenon). In our case of equal masses of two colliding
solitons, the classical and quantum descriptions lead to the same
result. Physically, the repulsive interaction of anti-phase matter
wave solitons can be regarded as exchange of velocities of two
colliding classical particles interacting via hard core potential.

\section{Numerical simulation of the two-soliton interactions}

In order to study the character of interaction between two
matter-wave solitons in numerical experiments we employ the idea
similar to that used for optical solitons in fibers
\cite{mitschke1987}. Initially at $t=0$, two solitons either
in-phase $\phi=0$ or out-of-phase $\phi=\pi$, are created at some
distance $\Delta_0$ from each-other. At later time $t = t_1 > 0$
the distance between solitons is measured again. If the solitons
did not interact, the distance between them should not change with
respect to its initial value $\Delta_1 = \Delta_0$. If the
interaction was attractive, the final measured distance should be
less than the initial distance $\Delta_1 < \Delta_0$. Finally
if the interaction was repulsive, the final distance should be
greater than the initial distance $\Delta_1 > \Delta_0$. The
numerical experiment consists in repeating the above procedure for
different values of the initial distance $\Delta_0$, starting from
sufficiently large separation, greatly exceeding the width of the
soliton then reaching small distances when solitons start to overlap.

\begin{figure}[htb]
\centerline{\includegraphics[width=6cm,height=6cm,clip]{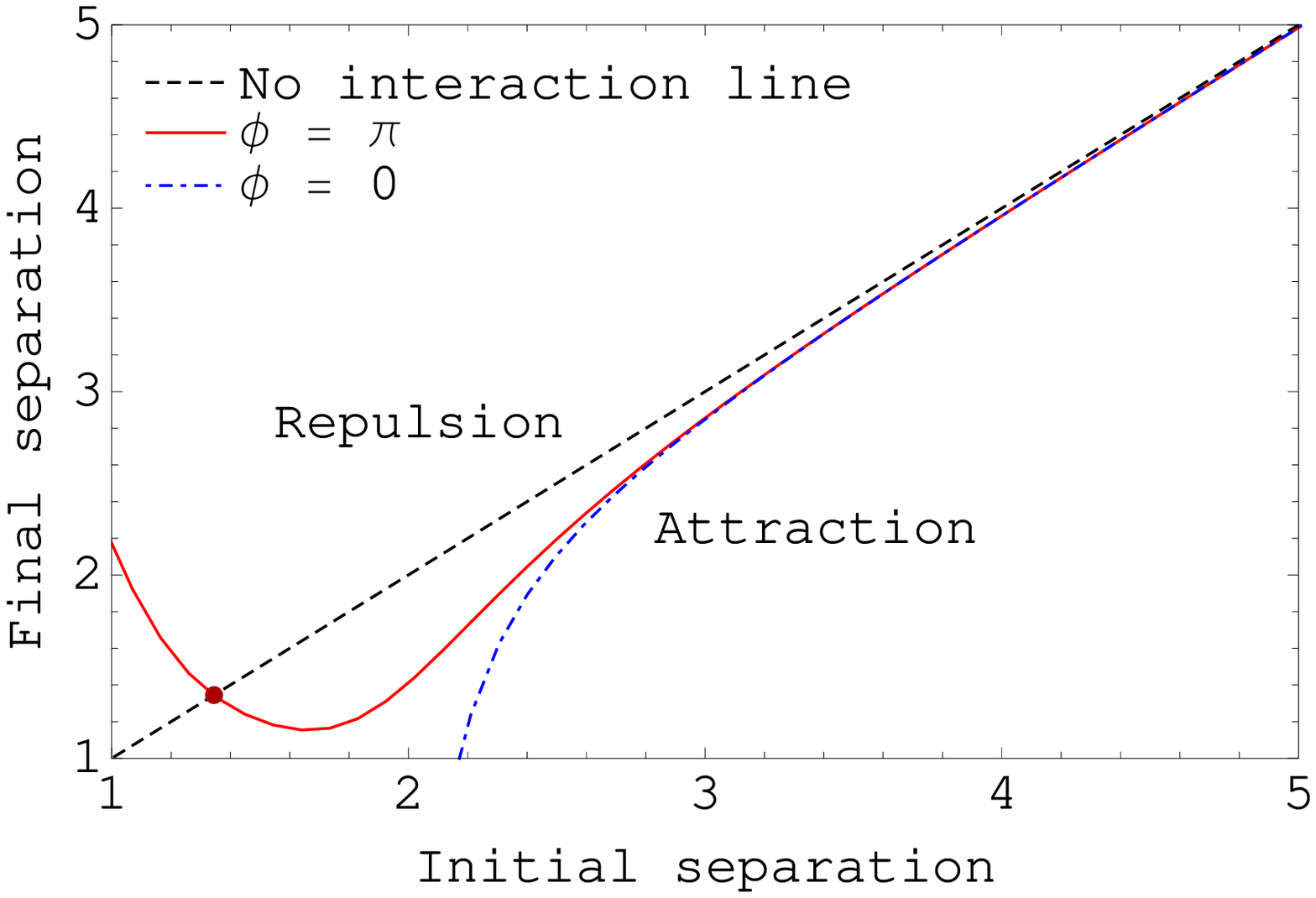}\quad
            \includegraphics[width=6cm,height=6cm,clip]{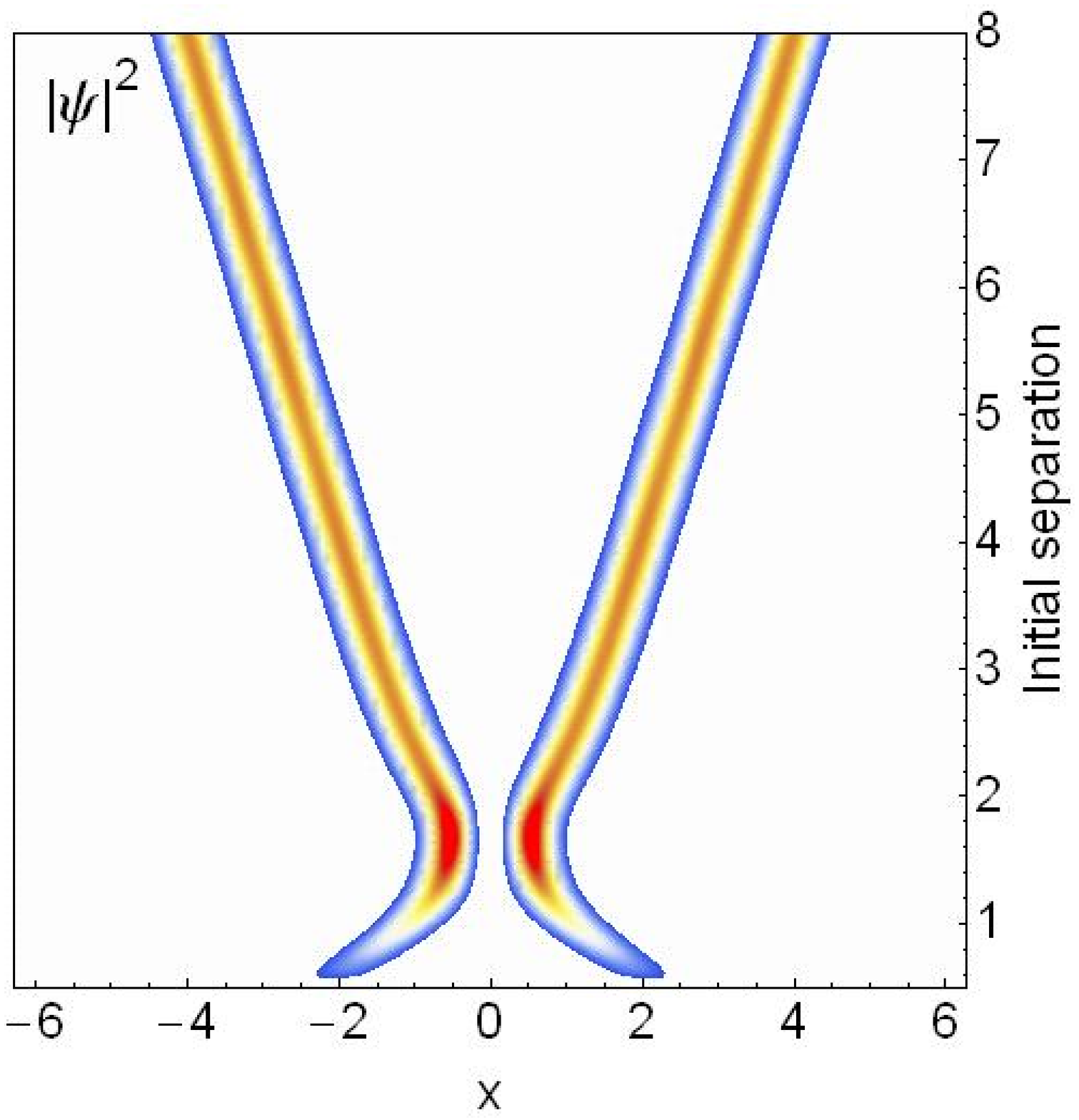}\quad
            \includegraphics[width=6cm,height=6cm,clip]{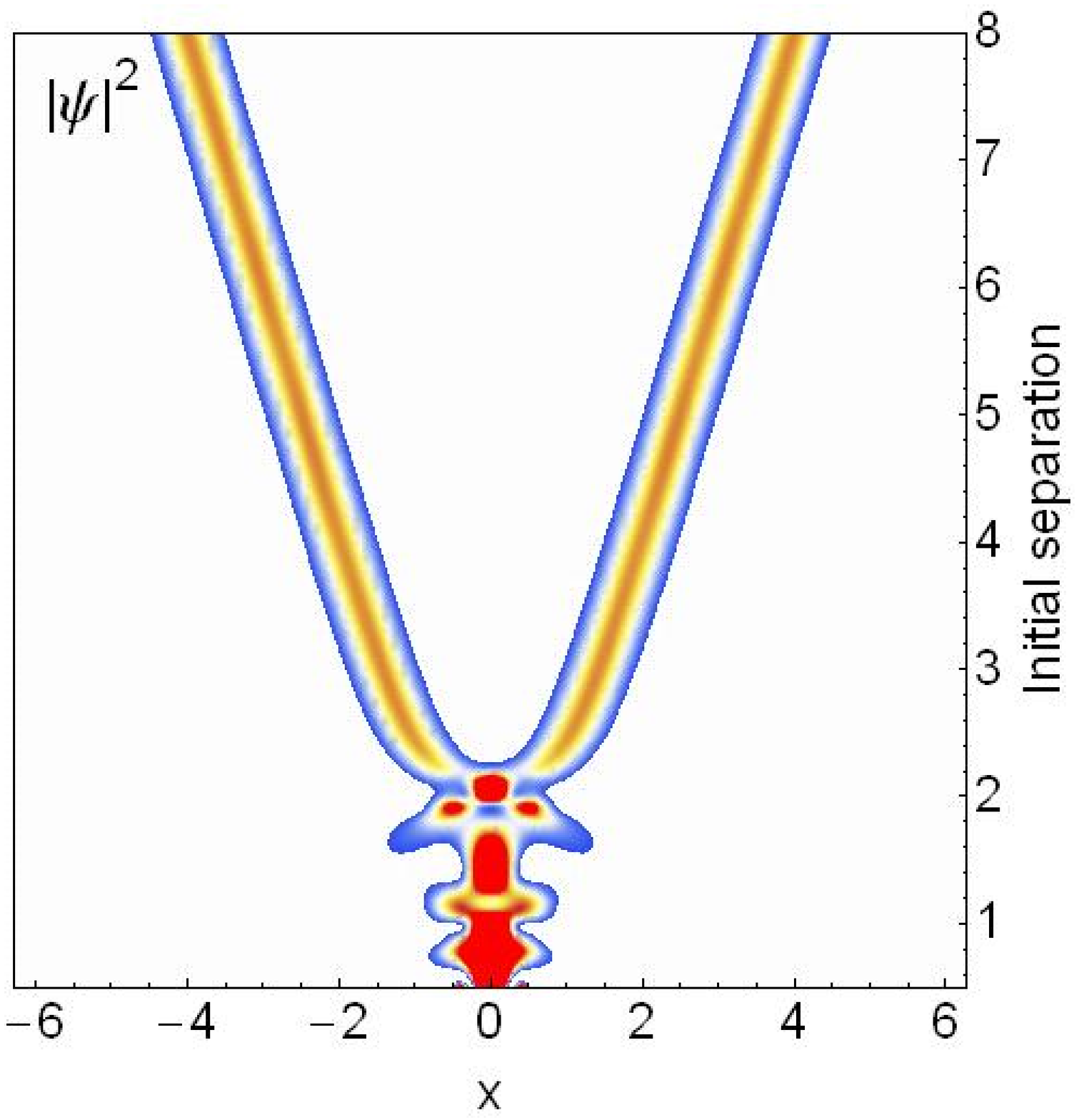}}
\caption{(Color online) Left panel: The character of soliton
interactions for anti-phase and in-phase solitons. Anti-phase
solitons attract each-other at large separations and repel at
small separations (red solid line). There exists a stationary
separation, shown by a red dot, where attraction changes to
repulsion. In-phase solitons always attract and collide (blue
dashed line). Middle panel: Anti-phase solitons can form a stable
soliton molecule at appropriate initial separation $\Delta_0
\simeq 1.3$. Right panel: In-phase solitons always collide and do
not form stable bound state.} \label{fig5}
\end{figure}

The result is presented in Fig. \ref{fig5} as a plot of $\Delta_1$
(final separation) vs. $\Delta_0$ (initial separation). When the
solitons, comprising the molecule, are out of phase ($\phi=\pi$)
we see that at large separations the solitons do not interact
($\Delta_1 \approx \Delta_0$, red curve is close to median), while
at small distances they attract each other ($\Delta_1 < \Delta_0$,
red curve is below the median) until they reach a minimum
(stationary) separation, where again $\Delta_1 = \Delta_0$ (red
dot on the median). When solitons are placed at even smaller
distances, they repel each other ($\Delta_1 > \Delta_0$, red curve
is above the median). That is, the two-soliton molecule behaves
like a diatomic molecule. For in-phase solitons ($\phi= 0$) we see
that solitons attract each other until their separation becomes
comparable to the width of the soliton, then merge forming a wave
packet, whose shape strongly oscillates.

The numerical simulations are performed using realistic values of
atom numbers and interaction parameters in $^{164}$Dy for which
$m=2.7 \times 10^{-25}$ kg, $d=10\,\mu_B = 9.27 \times 10^{-23}$ A
m$^2$, $a_{d}= \mu_0 d^2 m/(12 \pi \hbar^2) \simeq 7 \times
10^{-9}$ m. The frequency of radial confinement $\omega_{\bot} =
2\pi \times 62$ Hz provides radial oscillator length $l_0 \simeq 1
\, \mu$m and unit of time $t_0=2.6$~ms. For parameter values $g_0
= 20$ and $N=2$ used in numerical simulations we obtain the number
of atoms in a two-soliton molecule ${\cal N} = g_0 N l_0/(2 a_d)
\simeq 3000$. The total number of atoms in the $^{164}$Dy
condensate was ${\cal N} =15000$ \cite{lu2011}.

\section{conclusions}

We have studied the interaction between two bright solitons in a
dipolar BEC and found the conditions when they form a stable bound
state. It was revealed by numerical simulations of the governing
nonlocal GPE and corresponding variational analysis, that two
anti-phase  solitons in dipolar condensates behave similarly to a
diatomic molecule. Namely, they attract each-other at large
separation and repel each-other at small separation. There exists
a particular distance at which the two solitons remain motionless
in their stationary state. Solitons in a weakly perturbed molecule
perform small amplitude oscillations near the equilibrium
position, the frequency of which is predicted quite accurately by
the developed model. Two in-phase solitons when placed close to
each-other always collide and do not form the bound state. The
obtained results can be useful in further studies of the
properties of multi-soliton bound states in dipolar BEC.

\section*{Acknowledgements}
This work has been supported by King Fahd University of Petroleum
and Minerals under research group projects RG1333-1 and RG1333-2.
BBB thanks the Department of Physics at KFUPM and SCTP for the
hospitality during his visit.

\end{document}